\documentclass[sigconf,draft=true]{acmart}
\AtBeginDocument{%
  }

\usepackage{dirtytalk}
\usepackage{xcolor}

\newcommand{\kiev}[1]{\textcolor{red}{\textbf{Kiev: }#1}}
\newcommand{\alex}[1]{\textcolor{blue}{\textbf{Alex: }#1}}
\newcommand{\vicki}[1]{\textcolor{purple}{\textbf{Vicki: }#1}}
\newcommand{\filipe}[1]{\textcolor{green}{\textbf{Filipe: }#1}}
\newcommand{\vinicius}[1]{\textcolor{orange}{\textbf{Vinicius: }#1}}

\renewcommand{\kiev}[1]{}
\renewcommand{\alex}[1]{}
\renewcommand{\vicki}[1]{}
\renewcommand{\filipe}[1]{}
\renewcommand{\vinicius}[1]{}

\newcommand{\vibehack}{Vibe Hack}

\newcommand{\cscategory}{programmers}

\newcommand{\noncscategory}{non-programmers}

\setcopyright{none}
\renewcommand\footnotetextcopyrightpermission[1]{}

\graphicspath{ {./img/} }

\begin{document}

\title{"Can you feel the vibes?": An exploration of novice programmer engagement with vibe coding}

\author{Kiev Gama}
\affiliation{%
 \institution{Universidade Federal de Pernambuco}
 \country{Recife, Brazil}}
\email{kiev@cin.ufpe.br}

\author{Filipe Calegario}
\affiliation{%
 \institution{Universidade Federal de Pernambuco}
 \country{Recife, Brazil}}
\email{fcac@cin.ufpe.br}

\author{Victoria Jackson}
\affiliation{%
 \institution{University of Southhampton}
 \country{Southhampton, United Kingdom}}
\email{v.jackson@soton.ac.uk}

\author{Alexander Nolte}
\affiliation{%
 \institution{Eindhoven University of Technology, The Netherlands}
 \country{Carnegie Mellon University, USA}}
\email{a.u.nolte@tue.nl}

\author{Luiz Augusto Morais}
\affiliation{%
 \institution{Universidade Federal de Pernambuco}
 \country{Recife, Brazil}}
\email{gusto@cin.ufpe.br}

\author{Vinicius Garcia}
\affiliation{%
 \institution{Universidade Federal de Pernambuco}
 \country{Recife, Brazil}}
\email{vcg@cin.ufpe.br}


\begin{abstract}
Emerging alongside generative AI and the broader trend of AI-assisted coding, the term \say{vibe coding} refers to creating software via natural language prompts rather than direct code authorship. This approach promises to democratize software development, but its educational implications remain underexplored. This paper reports on a one-day educational hackathon investigating how novice programmers and mixed-experience teams engage with vibe coding. We organized an inclusive event at a Brazilian public university with 31 undergraduate participants from computing and non-computing disciplines, divided into nine teams. Through observations, an exit survey, and semi-structured interviews, we examined creative processes, tool usage patterns, collaboration dynamics, and learning outcomes. Findings reveal that vibe coding enabled rapid prototyping and cross-disciplinary collaboration, with participants developing prompt engineering skills and delivering functional demonstrations within time constraints. However, we observed premature convergence in ideation, uneven code quality requiring rework, and limited engagement with core software engineering practices. Teams adopted sophisticated workflows combining multiple AI tools in pipeline configurations, with human judgment remaining essential for critical refinement. The short format (9 hours) proved effective for confidence-building among newcomers while accommodating participants with limited availability. We conclude that vibe coding hackathons can serve as valuable low-stakes learning environments when coupled with explicit scaffolds for divergent thinking, critical evaluation of AI outputs, and realistic expectations about production quality.
\end{abstract}

\ccsdesc[500]{Applied computing~Interactive learning environments}

\keywords{Software Engineering Education, Hackathons, Vibe Coding, Generative AI}

\maketitle

\section{Introduction}
Software is a part of everyday life, from critical infrastructure and finance to health, mobility, and communication. Understanding how it functions and being able to contribute to its creation can thus be perceived as a form of civic participation~\cite{resnick2009growing}. Starting to learn about software is difficult, though, as it requires specialized knowledge that novices must acquire before being able to produce non-trivial artifacts~\cite{robins2003learning}. This expertise, in turn, is typically obtained through sustained education over longer periods of time, e.g., in the form of university courses. Sustained education, however, is not always feasible, particularly for individuals with limited time due to other commitments such as work and caregiving~\cite{kara2019challenges}.

The recent emergence of Generative AI (GenAI) might help lower this barrier. Vibe coding in particular appears to be promising~\cite{andrejkarpathy}. It allows basically anyone with access to suitable tools to create polished proofs-of-concept prototypes by simply describing what they want to build using natural language and iteratively refining what the tool produces~\cite{rooseNotCoderAI2025}. It is, however, unclear at this point if people actually learn about technology while vibe coding. What is clear, though, is that novices will unlikely benefit from trying vibe coding alone. Pedagogy has long established that learning in a social environment together with like minded individuals and suitable support and guidance greatly benefits learning~\cite{freeman2014active}.

Hackathons can be a suitable setting in this context. They are time-bounded events during which participants form teams to collaboratively develop an artifact, which often takes the form of a piece of software~\cite{falkFutureHackathonResearch2024}. Moreover, during hackathons, participants typically receive support in the form of mentors~\cite{nolte2020support}. Prior research has shown the suitability of hackathons for educational purposes both in formal~\cite{porras2018hackathons} and informal learning~\cite{nandi2016hackathons} settings. Hackathons have, however, also been criticized for being scary for newcomers to attend because of the assumption that technical proficiency is a requirement for attendance~\cite{gama2023developers,warner2017hack}.

It thus appears that running a vibe coding oriented hackathon can be mutually beneficial. Hackathons can provide a social environment where participants can try out and learn about technology while receiving support when needed and vibe coding can lower the barrier for attendance because it can lower the perception of technical proficiency being a prerequisite for attendance.


\filipe{Like this?}
In this paper, we report on an experience-based study of a one-day educational hackathon oriented toward vibe coding. Our goal was to understand how novices and mixed-experience teams utilize GenAI to transition from ideas to functional prototypes, what they learn in the process, and which event design choices facilitate inclusive participation.

\section{Background and Related Work}
In this section, we will introduce the concept of vibe coding before discussing how GenAI and hackathons have been utilized in the context of software engineering (SE) education.

\subsection{Vibe Coding}
Ongoing advances in AI, coupled with the natural language interfaces of AI-powered tools, such as Claude~\cite{claude}, Replit~\cite{replit}, Cursor~\cite{cursor}, and others, have led to the emergence of a new programming paradigm, colloquially known as \say{vibe coding}. First coined by the computer scientist Andrej Karpathy in a Tweet in February 2025, \say{vibe coding} is to \say{fully give in to the vibes, embrace exponentials, and forget that the code even exists...}~\cite{andrejkarpathy}. Popularly, vibe coding has been interpreted as developing software solely by prompting an AI with no technical knowledge or expertise required~\cite{oconnorVibeCodingNew2025, rooseNotCoderAI2025, kimWhatVibeCoding2025}.

Vibe coding has garnered significant attention within (e.g.,~\cite{smithAIVibeCoding2025, kimWhatVibeCoding2025}) and outside (e.g.~\cite{rooseNotCoderAI2025, talagalaWhatVibeCoding2025, stokel-walkerWhatVibeCoding2025}) the SE community, with some noting that it can democratize the development of new applications, as applications can be easily developed without requiring coding experience~\cite{rooseNotCoderAI2025}. Indeed, some consider vibe coding as a new development paradigm where humans and GenAI co-create software by conversing in natural language rather than code~\cite{meske2025vibecodingreconfigurationintent} or an evolution of first-generation AI-assisted programming~\cite{sarkarVibeCodingProgramming2025}, which started with the availability of tools such as Copilot. 
Benefits of orchestrating and prompting AI-based tools to create an application include a lower barrier to entry for application development~\cite{meske2025vibecodingreconfigurationintent}, reduced time to market~\cite {meske2025vibecodingreconfigurationintent}, productivity benefits~\cite{kimWhatVibeCoding2025}, and greater autonomy for engineers~\cite{kimWhatVibeCoding2025}. Yet, risks have also been identified, including the potential for developers' technical skills to degrade~\cite{meske2025vibecodingreconfigurationintent}, reduced code quality leading to issues such as security flaws~\cite{williamsWhatVibeCoding2025}, and ambiguous authorship and accountability for the code~\cite{meske2025vibecodingreconfigurationintent}. 

Despite the hype surrounding vibe coding, empirical research into this emerging phenomenon is scarce. One  study~\cite{geng2025exploringstudentaiinteractionsvibe}, with CS and SE students of differing coding experience, found that more inexperienced students interacted via prompts more than advanced students, who noticeably reviewed and amended code more frequently. Also, the prompting styles varied with inexperienced students authoring more vague prompts compared to the advanced students, who wrote more specific and code-focused prompts.

A study~\cite{sarkarVibeCodingProgramming2025} of vibe coding live-streams notes that professional developers use an ecosystem of AI tools, and the objectives and requirements for the application evolve through developers' interactions with the AI tools. Challenges faced include difficulties in communicating abstract ideas and understanding the capabilities and limitations of the AI tools. The study finds that technical expertise is beneficial in several ways, including identifying problems with the code, keeping the solution aligned with goals, and selecting the appropriate tools and technologies.

Finally, a qualitative study examining social media posts about vibe coding and data drawn from interviews of vibe coders~\cite{pimenovagoodvibrations2025} characterizes vibe coding as true co-creation between AI and humans. Vibe coders face challenges, including incorrect solutions provided by the AI and poor-quality code. Best practices to overcome these specific challenges include breaking up large tasks and rubber-ducking.

By setting our experience report in a different context to these prior studies, we are able to contribute an initial understanding of the impacts of vibe coding in a context important to SE education. Namely, a time-bound educational hackathon open to all students irrespective of technical expertise and degree studied. In such a setting, we can extend the above findings to include the perspectives of non-developers and contribute findings pertinent to hackathons.

\subsection{GenAI in SE Education}
With the rapid adoption of GenAI tools by software professionals~\cite{stackoverflow_dev_survey_2025}, computer science educators are integrating GenAI into higher education curricula~\cite{pratherBeyondTheHype2025} to ensure that students acquire the AI literacy and skills to thrive in the software development profession. GenAI tools have been incorporated into programming classes (e.g., ~\cite{prather2023s, pratherWideningGap2024, keuningStudentPerceptions, kerslakePrompting2024}), and other aspects of the development process, including testing~\cite{tuzmenUseVV2024} and software design~\cite{camaraGenerativeModeling2024}. 

On the one hand, integrating GenAI in classes offers benefits to students such as accelerating task progress~\cite{pratherWideningGap2024} and increasing their self-efficacy~\cite{yilmazGenAIEfficacy2023}. On the other hand, educators have highlighted drawbacks with using GenAI tools, including an over-reliance when programming~\cite{prather2023s}, meta-cognitive difficulties~\cite{pratherWideningGap2024}, and eroding social interactions with peers~\cite{houEroding2025}. Despite these challenges, GenAI has become a core component of modern software development, and it is essential for students to understand how to interact with GenAI tools effectively~\cite{kerslakePrompting2024}, its capabilities and the risks it poses, such as introducing bias~\cite{huangbias2024} and presenting ethical and legal~\cite{alkamliEthicalLLM2024} concerns. Exactly how to provide such learning is a challenge for educators~\cite{petrovskaIncorporateGenAI2024}. One such approach, as described in this paper, is to run a hackathon, a popular approach for complementing traditional classroom learning~\cite{gamaHackathonsFormalLearning2018}. 

\subsection{Hackathons and SE Education}

Hackathons are time-bounded events during which participants form teams and collaboratively work on a project that is of interest to them~\cite{falkFutureHackathonResearch2024}. The outcome of the project commonly takes the form of a software prototype. Hackathons have been adopted in various domains including science~\cite{nolte2020support}, industry~\cite{pe2022corporate}, entrepreneurship~\cite{nolte2019touched}, government~\cite{johnson2014civic}, civic engagement~\cite{yuan2021open}, and others. One popular form of such events are educational hackathons. These can come in the form of informal learning spaces~\cite{nandi2016hackathons} or they can be integrated into formal learning complementing what occurs in classrooms~\cite{porras2018hackathons,willis2017challenge}. Hackathons provide an excellent environment for authentic learning (i.e., solving problems in realistic contexts that ensure relevance)~\cite{hogan2022hackathons}. The adoption of hackathons in education has been steadily expanding~\cite{falk202010} and being applied in courses across multiple disciplines (e.g., software engineering\cite{porras2018hackathons}, security~\cite{amefon2022}, IoT~\cite{gamaHackathonsFormalLearning2018}).

The short duration of hackathons constrains the exploration of software engineering practices (e.g., design, testing)~\cite{gama2017preliminary,steglich2020hackathons}. There have been attempts to explore specific SE practices, such as code reviews~\cite{paganini2020engaging,crusoe2016channeling}. Nevertheless, SE learning and knowledge sharing in informal hackathon contexts within higher education~\cite{steglich2021online} typically covers broader topics and occur through peer interaction and situated practice, providing learning opportunities for newcomers to specific domains~\cite{nolte2020support}.


Although research on educational hackathons has highlighted their potential, several challenges to their inclusivity remain. One major issue is the lack of diversity, with events often attracting skilled developers, which can be intimidating for individuals with limited programming experience~\cite{gama2023developers}. While organizers have experimented with making events \say{beginner-friendly} by providing mentor support~\cite{nolte2020support}, these interventions often steer novices away from core development tasks~\cite{taylor2018everybody}. Compounding this issue, the significant time commitment of traditional multi-day hackathons not only excludes participants with personal or professional obligations but also intensifies the pressure and potential for fatigue, especially for newcomers~\cite{gamaHackathonsFormalLearning2018}. Therefore, designing shorter, more focused events emerges as a key strategy to foster a more welcoming and accessible environment. The 8-hour format of the hackathon we organized as described below, was a deliberate choice to address these dual challenges of intimidation and exclusion, creating a space where participants with limited availability could meaningfully engage with software development.

\section{Setting}
The vibe coding hackathon was conceived as an in-person and inclusive event designed to welcome early-stage undergraduate students with diverse backgrounds (Computer Science, Cinema, Engineering, Design, etc) and levels of experience in programming, ranging from complete beginners to advanced programmers. Its main goal was to foster a collaborative environment where learning and enjoyment were as central as the technical outcomes. To encourage this spirit, teams were organized into two categories: those composed exclusively of undergraduate students from computing programs (Computer Science, Computer Engineering, Information Systems, and Artificial Intelligence), and mixed teams integrating participants from both within and outside the computer science domain. This structure enabled cross-disciplinary collaboration while also acknowledging the value of peer exchange among students at various stages of their learning journey.

The event, held at a Brazilian public university, received financial support from two companies: a multinational that specializes in building information systems and a local software house.

\subsection{Key Event Design Decisions}
To facilitate replication, we structure this section according to the Hackathon Planning Kit~\cite{affia2025organize}, which served as a guide for organizing this event. The 12 key decisions taken in this hackathon were as follows:
\begin{enumerate}
    \item \textbf{Goal}: The hackathon’s goal was to practice ``vibe coding'' while fostering an inclusive, collaborative learning environment open to beginners and experienced students across disciplines. It aimed to spark creativity, form small teams, and connect participants with industry sponsors.
    
    \item \textbf{Theme}: The organizers developed a \say{theme generator app} that would create a sentence defining the challenge (e.g., how to make learning more inclusive for Black people, considering floods). It would work as an example of vibe coding and also support the team in generating a theme/problem to be addressed in their project. \vicki{Do we need to explain why we took this approach of a theme generator, rather than the organizers stating a theme? Moreover, if we have space, perhaps we should add a screenshot of the theme generator?}
    
    \item \textbf{Competition vs. Cooperation}: A cooperative atmosphere was fostered under a light competitive layer to recognize standout work without discouraging novices. The panel of judges consisted of three people from industry, two of them with experience in vibe coding. Two projects were awarded for technical quality, innovation, and social relevance. Winners of each category received 3D-printed trophies.
    
    \item \textbf{Stakeholder Involvement}: The topic was broad and no specific stakeholder was in mind. One of the sponsoring companies uses vibe coding to demonstrate concepts in some of their projects. Two people from their personnel were in the event and had brief discussions with some of the teams. 
    
    \item \textbf{Participant Recruitment}: The hackathon was announced by the Informatics Department press office through its social networks. The online registration form was closed one week after the announcement, when it surpassed 100 registrations. Due to physical space limitations of the room, 50 participants were selected but only 31 showed up.
    
    \item \textbf{Specialized Preparation}: During the general instructions there was a quick contextualization about vibe coding and some tools. The theme generator app was also presented as an example of vibe coding. The prompt is available online~\cite{supplementarymaterial}.
    
    \item \textbf{Duration \& Breaks}: Since vibe coding generates executable results faster, we chose a one-day hackathon format (approximately 9 hours). Just one break was scheduled, with coffee, refreshments and snacks sponsored by a local IT company.
    
    \item \textbf{Ideation}: Participants had slightly more than one hour to submit their project idea that would be vibe coded afterwards. Their brainstorming process was supported by the theme generator app. They were also free to use LLMs to support them in that process.
    
    \item \textbf{Team Formation}: Teams consisted of 3 or 4 individuals and two categories: \cscategory and \noncscategory. Participants could have formed their teams previously. At the beginning of the hackathon, the schedule included dedicated time for team formation and onboarding.
    \item \textbf{Agenda}:
    
\noindent{\raggedright
08:30 \textendash\ Registration\\
09:00 \textendash\ General instructions and challenge briefing\\
09:45 \textendash\ Team formation (for those without teams)\\
10:00 \textendash\ Ideation begins\\
11:00 \textendash\ Submit project idea and start development\\
13:00 \textendash\ Quick Snack/Coffee Break\\
16:00 \textendash\ Pitches begin (final presentations)\\
17:30 \textendash\ Awards and Closing\par}

    \item \textbf{Mentoring}: The mentoring team consisted of three students and three teachers. Guidance was mainly about overall instructions, checkpoints and event clarifications. During the hackathon, participants demanded no technical support.
    
    \item \textbf{Continuity Planning}: There were no plans on the continuation of solutions.

\end{enumerate}

\subsection{Winning Projects}
There were 31 participants split into 9 teams (4 of them cross-disciplinary). Among the projects presented, two stood out as winners and were recognized for their technical quality, innovative approach, and social relevance. In the mixed-team category, the winning project was a gamified web platform designed to promote inclusion and accessibility in recruitment processes for neurodivergent individuals. The solution leverages artificial intelligence to adapt the selection journey to the user’s profile, considering, for instance, candidates with Attention Deficit Hyperactivity Disorder (ADHD). Traditional assessment stages are reimagined as short missions with accessible design, simplified text, audio resources, and adjustable pacing, aiming to reduce cognitive overload and enhance engagement.

In the category restricted to CS-related students, the winning project was IRIS Map, an interactive web platform focused on fostering safety, visibility, and community support for LGBTQIAP+ individuals. The system uses maps to highlight safe spaces, businesses led by community members or allies, and inclusive events. By strengthening support networks and making diversity more visible, the initiative seeks to contribute to building more inclusive and respectful environments through technology.

\section{Method}
To better understand the participants' experience of \vibehack, we decided upon three research activities. The first was observations of teams made during the event itself, the second was an (optional) exit survey completed by participants at the end of the Vibe Hack, and the third was to invite participants for an interview. Having three different perspectives on the experience would enable us to gain richer insights into it. Data collection instruments are available on FigShare~\cite{supplementarymaterial}.

\begin{table}[]
\caption{Interviewees profile}
\begin{tabular}{lclll}
\toprule
\textbf{ID}  & \textbf{Gender} & \textbf{Bachelor} & \textbf{Team}   & \textbf{Team Category}      \\
\midrule
P01  & M      & Info. Systems       & Team 1 & Tech-only          \\
P02  & M      & Info. Systems       & Team 1 & Tech-only          \\
P03  & M      & Info. Systems       & Team 2 & Cross-disciplinary \\
P04  & M      & Comp. Science       & Team 3 & Tech-only          \\
P05  & F      & Comp. Science       & Team 4 & Cross-disciplinary \\
P06  & F      & Info. Systems       & Team 2 & Cross-disciplinary \\
P07  & F      & Design   & Team 2 & Cross-disciplinary \\
P08  & M      & Comp. Eng.      & Team 5 & Tech-only          \\
P09  & M      & Film/Cinema   & Team 6 & Cross-disciplinary \\
P10 & F      & Lang. \& Literature       & Team 6 & Cross-disciplinary \\
\bottomrule
\end{tabular}
\label{tab:interviewees}
\end{table}

\subsection{Observations During the Hackathon}
We conducted non-participant observations throughout the hackathon. Three teachers served as observers and held brief discussions during the event to align focus and compare emerging impressions. One observer adopted a roaming role, circulating among teams, staying several minutes with each to watch interactions, and recording descriptive field notes. The observers periodically shared their observations to clarify interpretations and consolidate salient episodes, providing informal triangulation across vantage points.

\subsection{Exit Survey}
An optional exit survey was prepared to gather feedback from participants about their experience of the hackathon.

\subsubsection{Survey Instrument}
Specific focus areas within the instrument were on the teams’ creative process including their usage of GenAI, quality of the final solution, incorporation of architecture considerations into the design, and overall learnings from the experience. Originally, a goal of the survey was to analyze students software engineering self-efficacy framed by Bloom's taxonomy~\cite{krathwohlbloomrevision2002}. On reflection, inserting such questions would have led to a long survey, introducing the risk of low completion rates. As it was decided more important to increase the survey completion rate than explore self-efficacy, the majority of the SE self-efficacy questions were cut from the survey, with the exception of the architecture questions. 
The survey consisted of 50 questions in total with the majority (44) of the questions written as mandatory, closed-ended with extensive use of Likert scale questions (31) with only 6 optional, open-ended questions.~\autoref{tab:survey} summarizes the questions asked in each focus area. The full set of questions is available~\cite{supplementarymaterial}.

Several of these questions probed not only whether teams generated original or varied ideas, but also how frequently they revisited or abandoned them, and whether GenAI influenced the extent of iteration. Open-ended items further captured participants’ descriptions of their ideation stages, allowing us to examine convergence or divergence in creative exploration.

The survey was conducted in Portuguese and accessible anonymously to the hackathon participants via Google Forms. No personal identifiable information, including demographics, was collected. The survey first described the purpose of the research before asking for informed consent to participate in the survey. Participants were invited to complete the survey at the end of the hackathon. Of the 31 participants, 27 completed the exit survey fully. This data was subsequently analyzed and presented in~\autoref{survey_results}.

\begin{table}
    \centering
    \caption{Areas covered by exit survey questions}
    \begin{tabular}{p{0.18\linewidth}p{0.06\linewidth}p{0.65\linewidth}}
    \toprule
        \textbf{Area} & \textbf{Num.} & \textbf{Questions}\\
        \midrule
         Motivation /Experience& 14 & Studied subject, programming experience, familiarity with GenAI tools, hackathon exp., motivations for attending\\
         Creative Process& 13 & Use of AI for ideating, interaction with AI, description of creative process\\        
         Team Process& 6 & Roles on team, collaboration effectiveness and engagement \\
         Quality& 5 & Quality of AI responses and final solution\\
         Architecture&7 & Architecture self-efficacy based on Bloom's taxonomy \\
         Learnings & 5 & specific skills learned, intended use of AI going forward for developing applications \\
         Feedback& 1& Feedback on Vibe Hack\\
         \bottomrule
    \end{tabular}
    \label{tab:survey}
\end{table}

\subsubsection{Data Analysis}
The survey responses were captured automatically into a Google Sheet and subsequently translated into English using Google Translate. A bi-lingual researcher fluent in Portuguese and English corrected any translation errors. In addition to quantitative summaries, we occasionally report anonymized verbatim excerpts from the survey’s open-ended fields when they help contextualize the findings. All Portuguese excerpts were translated into English by a bilingual researcher; minor edits were applied for clarity while preserving the original meaning.

Next, the data was analyzed within Google Sheets. Basic counting was performed on the responses as presented in~\autoref{survey_results}.

\subsubsection{Survey Respondents' Experience} \label{prior_experience}
The respondents' prior experiences of hackathons, GenAI, coding, and software architecture are summarized below:
\vicki{do not know if these should mention how the non-technical participants responded such as the 4(?) non-CS not having never heard of software architecture and no coding experience}
\begin{itemize}
    \item \textbf{Hackathon experience:} For 20 respondents, this was their first hackathon. 4 had attended one hackathon, 1 had attended two hackathons, and 2 respondents had attended more than three.
    \item \textbf{GenAI familiarity:} The majority (23) of the respondents were already familiar with GenAI with 9 having moderate familiarity and 14 having high familiarity levels. Only 4 had little familiarity.
    \item \textbf{Coding experience:} Compared to peers, most (19) rated their coding experience as comparable (12) or more experienced (4). In contrast, 3 felt they were very inexperienced and 8 felt inexperienced.
    \item \textbf{Knowledge of software architecture:} Respondents generally had low levels of knowledge (20) with 2 having never heard of it, 11 having heard about it but no technical knowledge, and 7 claiming basic knowledge. Only 6 claimed intermediate and 1 with advanced knowledge. 
\end{itemize}

In terms of disciplinary background, the respondent pool was predominantly from computing-related programs (Information Systems, Computer Science, Computer Engineering). However, four respondents came from non-computing majors (e.g., Biomedicine, Design, Film/Cinema, and Languages \& Literature). While numerically in the minority, these individuals offered qualitatively distinct perspectives during ideation and teamwork, which we elaborate in Section \ref{survey_results}.

\subsection{Interviews}
While the surveys provided breadth and quantitative insights into participants' experiences, we wished to probe further into individual experiences by conducting interviews. We purposefully sought diversity in experiences by selecting participants who differed in terms of their gender, degree subject, and team category (tech-only or cross-disciplinary). We found ten participants who were willing to be interviewed (\autoref{tab:interviewees}) coming from five different teams. Six were identified as men, four as women. Three studied non-technical subjects, such as Design. Six came from cross-disciplinary teams.

Interviews were conducted by three researchers. They were conducted in Portuguese and remotely via Google Meet the week immediately after the event, to minimize recall bias and the telescoping effect while the experiences were still salient. We used a semi-structured format following an interview protocol \cite{supplementarymaterial}. The interview covered the usage of generative AI in creative and technical software tasks: prior exposure, idea generation, how AI was used (tools, prompts, leadership), usefulness and reliability, and technical understanding. It also inquired about challenges, team dynamics, personal reflections, learning, future use, and concluded with an open-ended question.


\subsubsection{Data Analysis}

\filipe{Please, check if it has enough information:}

Audio recordings were transcribed in Portuguese using an automatic speech recognition service based on OpenAI Whisper version 3. Transcripts were anonymized by replacing participant names with identifiers P01 to P10. One researcher then reviewed the transcripts to correct recognition errors, with particular attention to proper nouns and tool names that were frequently mistranscribed, such as Lovable, ChatGPT, and Claude.

Analytic coding proceeded in two stages. First, two researchers divided the transcripts, with each researcher taking a separate portion, and selected relevant excerpts from their assigned sections. They attached open, inductively generated codes to each excerpt, and there were no intersections of quotes evaluated by both. This process produced a dataset of 294 quotes, each paired with its associated code comments. Second, the table containing quotes and codes was provided to ChatGPT running GPT-5 Thinking to assist with code organization. The prompt was enhanced using the OpenAI GPT-5 Prompt Optimizer\footnote{\url{https://platform.openai.com/chat/edit?models=gpt-5&optimize=true}} and can be found in the Supplementary Material ~\cite{supplementarymaterial}. This model-assisted approach was chosen both to enhance the efficiency of synthesizing the codes and to serve as a check against researcher bias by surfacing thematic clusters from the ground up. The prompt required clustering of the existing code comments and mandated that each proposed cluster be explicitly supported by the related quotes. 

The use of LLMs in qualitative research in Software Engineering is discussed and encouraged to automate and streamline certain phases of research, but the decision-making should remain as an essential part of human participation ~\cite{trinkenreich2025get,bano2024large}. In this experience report, relevant quotes were selected by two researchers, with the assistance of LLM tools to support the analysis. We recognize that LLMs cannot replace human interpretive judgment in qualitative analysis. Therefore, all clustering outputs were treated as organizational suggestions rather than definitive findings.


\vinicius{Don't you think we need to add a sentence here describing how the two researchers ensured coding consistency? A suggestion -- as a comment -- follows}
\vinicius{One question (but I think it might not be worth the time/effort) is whether it's worth calculating IRR using Cohen's Kappa?} 
\vinicius{Let's add a brief rationale for *why* we chose to use an LLM here :D}

\kiev{It is important to provide the prompt(s) in the supplementary material}

To prepare results for reporting, the selected Portuguese quotes were translated into English. The complete analysis, including automated clustering, was subjected to human verification by a researcher who examined the coherence of clusters, checked quote assignments, and confirmed that no hallucinated content or incorrect groupings had been introduced. Model-assisted steps were therefore used as an organizational aid, with final interpretive decisions grounded in the researchers’ review of the original language data and corresponding translations.

\subsection{Ethics and Data Availability}
The study was approved by the IRB of the institution hosting \vibehack. Informed consent was received from all survey and interview participants. No monetary or other incentives were provided for participating in the research study.

The IRB project made explicit that the participant data would not be fully shared, but only excerpts would be used to report the findings. The survey instrument and interview protocol are available~\cite{supplementarymaterial}.

\vicki{What about releasing the code for the theme generator?}
\vicki{and from a timing perspective, no one reviews a paper the day it is submitted, so we have a few days past the deadline to tidy up any data in Figshare.}
 
\section{Results}
We first present the results from observations taken during \vibehack, followed by the exit survey results, and the findings from the ten interviews.

\subsection{Observations}
Our analysis of hackathon teams using vibe coding revealed distinct patterns in collaboration, tool adoption, and encountered challenges. 

Teams with mixed backgrounds tended to combine brainstorming, exploratory prototyping, and iterative refinement, often leveraging tools such as ChatGPT, Gemini, and Lovable to generate ideas, validate directions, and create interactive front-ends. Computer science–only teams, in contrast, relied on generated code as a starting point for building their applications, which they then refined by manually fixing bugs and enhancing functionality; in some cases, they also utilized generative AI to produce user interface elements, such as app icons. 

Across groups, vibe coding was used to accelerate development through tools like Copilot, VSCode integrations, and Lovable interfaces. However, participants highlighted recurring obstacles, including limited credits to test prompts, difficulties in connecting disparate tools, and the challenge of manually editing or scaling prompts. Vibe coding practices enabled rapid prototyping and cross-disciplinary collaboration. However, tool fragmentation and resource constraints limited their full potential, shaping both the pace and direction of project outcomes.

As another observation, we noticed that all vibe-coded web applications tended to have very similar user interfaces, lacking much innovative design or a combination of widgets. This also happened with the theme generator app. Two teachers utilized distinct tools (Replit and Loveable) to generate two similar apps, but with different prompts that targeted the same objective. 

\subsection{Survey} \label{survey_results}
This section presents the survey results with each section corresponding to the different question groupings in the exit survey described above.

\subsubsection{Motivation/Experience.} 
\begin{figure}
    \centering
    \includegraphics[width=0.95\linewidth]{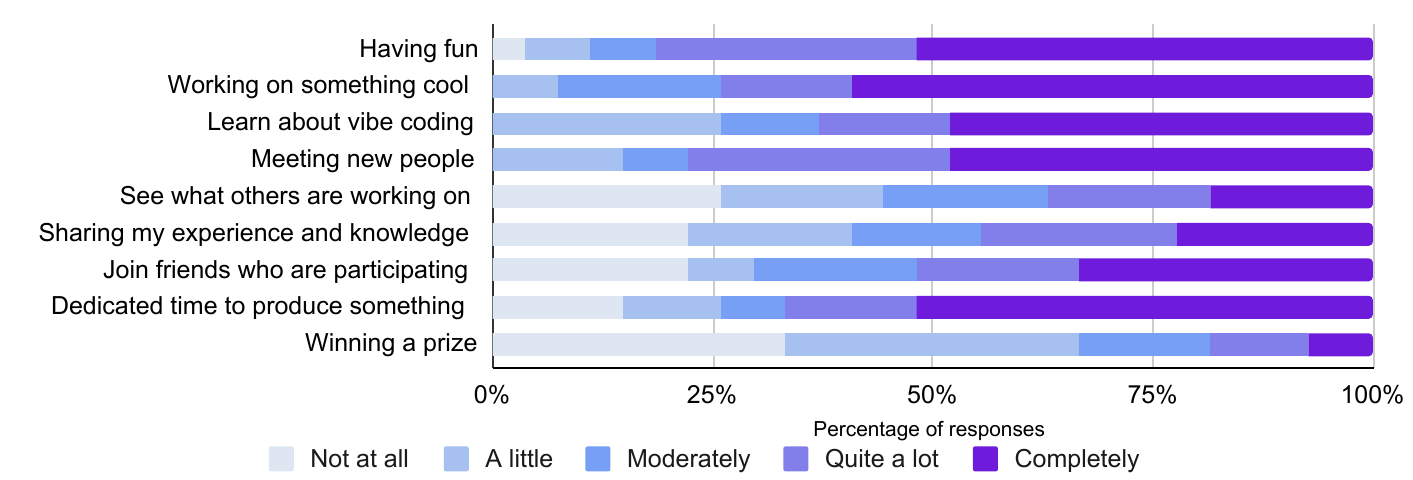}
    \caption{Respondents' motivations for attending}
    \Description[Nine reasons for participating.]{Nine reasons for participating: having fun, working on something cool, learn about vibe coding, meeting new people, see what others are working on, sharing experience and knowledge, joining friends who are participating, dedicating time to produce something, winning a prize. Over half of the respondents strongly agreed that they were participating to have fun, work on something cool, learn about vibe coding, meet new people, and have dedicated time to produce something. Only a handful of people were motivated by the prospect of winning a prize.}
    \label{fig:motivations}
\end{figure}

Our respondents were predominantly motivated to participate for social reasons with having fun, working on something cool, meeting new people, and joining friends who are participating all rated as positive motivations for attending (see~\autoref{fig:motivations}). Learning in an environment where they had \say{Dedicated time to produce something} were both important motivating factors also, with nearly half \say{Completely} agreeing that \say{learning about vibe coding} was a motivation for participating. Winning a prize was the least important motivation. These motivations re-enforce the view~\cite{porras2018hackathons} that hackathons are fun, social events providing an opportunity to learn new technologies. Beyond the Likert-scale items, open-ended responses also suggest a confidence-building role for the event among first-timers. One anonymous respondent explicitly framed the hackathon as a gentle on-ramp: \say{\textit{I want to start participating in these events, but I’m new to this world … I hope to gain confidence to go to other kinds of hackathons}} (exit survey, open-ended motivation; translated from Portuguese; ellipsis in original). While only one out of the four “other motivations” entries mentioned confidence explicitly (out of 27 total respondents), the comment points to a perceived reduction in the psychological barrier to entry for novices. 
\vinicius{I tried to highlight the security-related motivations for starting to experiment with AI-supported hackathons (I left the previous section commented. The new one starts with "Beyond the Likert-scale...")}

Taken together with the predominantly social/learning motivations in Figure \ref{fig:motivations}, this suggests that short, inclusive, time-bounded formats can function as confidence-building environments for newcomers—complementary to, rather than substitutive of, traditional coursework.

It is worth noting that the diversity of disciplinary background was limited: only four out of 27 respondents came from non-computing programs. Yet, even in this small group, we observed contributions that broadened the project's framing. For example, survey and interview data indicate that design students made decisions about user experience, while audiovisual and biomedicine students contributed to narrative framing and accessibility considerations. Although not sufficient to claim representativeness, these accounts illustrate how cross-disciplinary presence can enrich vibe coding teams.

\subsubsection{Creative Process.} 

\begin{figure}
    \centering
    \includegraphics[width=0.8\linewidth]{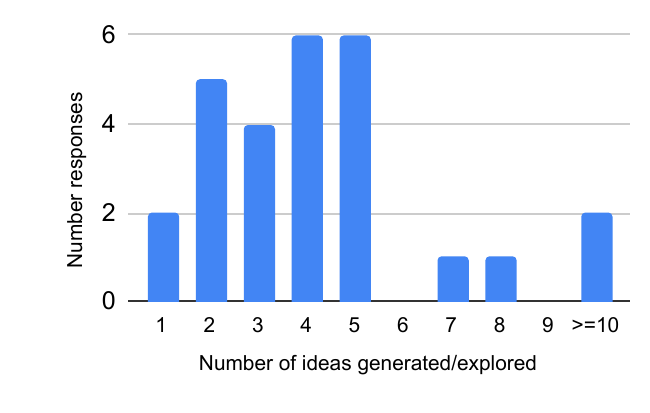}
    \caption{Number of ideas generated}
    \label{fig:num_ideas_generated}
    \Description[Number of ideas generated]{Number of responses against number of ideas generated/explored. 2 generated 1 idea, 5 generated 2 ideas, 4 generated 3 ideas, 6 generated 4 ideas, 6 generated 5 ideas, 1 generated 7 ideas, 1 generated 8 ideas, and 2 generated ten or more ideas.}
\end{figure}

At the start of the hackathon, each team needed to define a problem they wished to solve, identify and explore different ideas that could help, and to select one to be developed further, preferably into working code. As the teams were \say{vibe coding}, we were curious to understand their creative process and the involvement of AI tools. Firstly, as shown in~\autoref{fig:num_ideas_generated}, the number of ideas considered and explored was five or fewer, as reported by the majority of individuals (23/27 respondents). This is perhaps not surprising given the time-bound nature of the event and the focus on delivering working code. In contrast, two respondents noted their team explored more than 10 ideas, with one team generating 15 ideas. 

\begin{figure}
    \centering
    \includegraphics[width=0.95\linewidth]{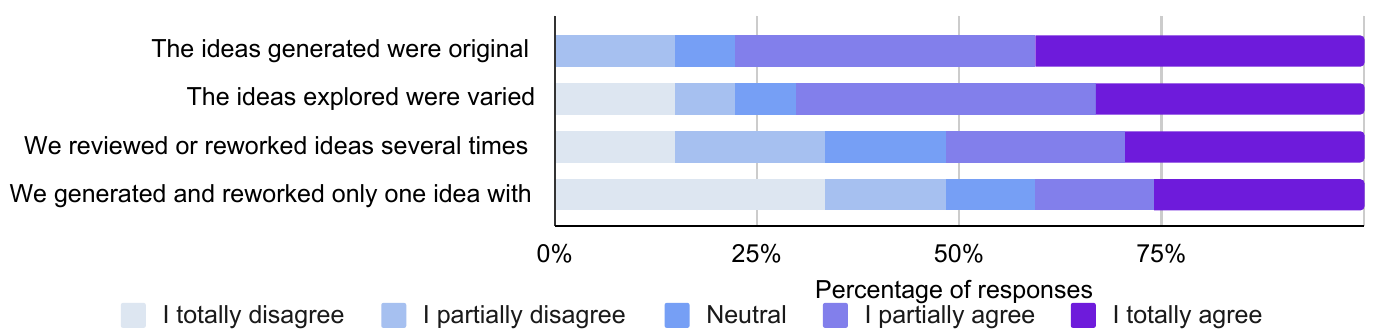}
    \caption{Exploring ideas with GenAI}    
    \label{fig:ideation}
    \Description[Exploring ideas with GenAI]{Shows percentages of responses that totally disagree, partially disagree, neutral, partly agree, and totally agree against four statements. The first was the ideas generated were original, the second was the ideas explored were varied, the third was we reviewed or reworked ideas several times, and the fourth was We presented and reworked only one idea with no different alternatives.}
\end{figure}

Secondly, about the use of GenAI in the ideation process (\autoref{fig:ideation}), over 75\% of the respondents partially or totally agreed that the ideas generated were original, with slightly fewer (70\%) feeling that the ideas explored were varied. These responses demonstrate a broad, optimistic view of GenAI's ability to assist with ideation through its capacity to generate original and diverse ideas. Yet, there was some evidence that ideas were little reviewed or reworked, with just over half of the respondents partially or totally agreeing that ideas were reviewed or revised, and 40\% partially or totally agreeing that they only generated and reworked a single idea. 


Finally, in terms of guiding the creative process, all teams used AI to assist, yet humans were predominantly the driving force. Ten respondents noted that humans were driving, with some help from AI, 9 felt it was balanced between AI and humans, and 5 felt it was mostly humans with little help from AI. In contrast, only 3 answers said that AI contributed a lot with little human involvement.

\subsubsection{Team Process.} 
\begin{figure}
    \centering
    \includegraphics[width=1\linewidth]{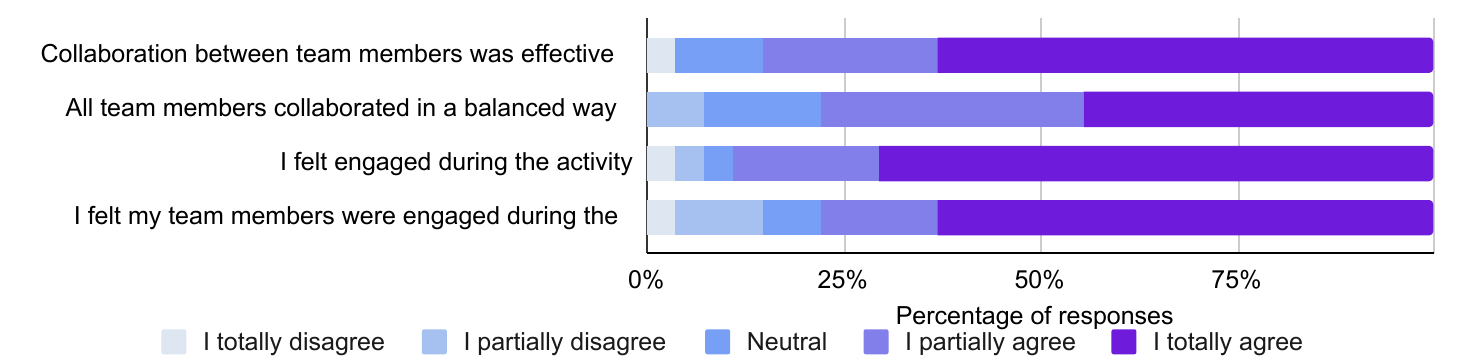}
    \caption{Team collaboration effectiveness}
    \label{fig:collaboration}
    \Description[Team collaboration effectiveness]{Shows percentages of responses that totally disagree, partially disagree, neutral, partly agree, and totally agree against four statements. The first was collaboration between team members was effective, the second was all team members collaborated in a balanced way, the third was I felt engaged during the activity, the fourth was I felt my team members were engaged during the activity.}
\end{figure}

The majority of respondents were very positive about the effectiveness of their team's collaboration and individuals' engagement in the activity (\autoref{fig:collaboration}). More than 75\% of respondents felt collaboration between team members was effective (partially or totally agree) and that all team members collaborated evenly. Moreover, nearly 90\% felt engaged during the activity, and more than 75\% felt their team members were also engaged.

To work effectively, the individuals in the teams needed to self-organize with some teams deciding to allocate specific roles to individuals. Only 11 respondents noted their team had a designated leader. Alongside this leader role, 14 respondents indicated there was a lead programmer on their team, 12 indicated a designer role was present, while 9 indicated less role formality in their team. 

In terms of who in the team interacted with the GenAI tools, there are two distinct interaction patterns. In the first, there is a single GenAI instance with either the whole team sharing it (3 respondents) or one person in the team interacting with it (2 responses). In the second, multiple GenAI instances were used, with 7 respondents noting all team members used their own instances and 15 indicating they used multiple instances to compare responses. 

For the majority of respondents (16), their interactions with the AI tools took place continuously throughout the hackathon. Three respondents noted it was mostly used at the beginning, 6 indicated it was mostly used in the middle, and 2 felt there was no clear interaction pattern.

Finally, a significant proportion of respondents (66\%) felt that using AI in the hackathon had freed up time.
\vicki{I am puzzled by the freetext survey question related to how this time saving was used. Do the answers refer to the activities AI sped up or, because AI did some work and thus saved time, these were activities they could spend more time on without AI. Not quite sure how to interpret the responses.}

\subsubsection{Quality.} 
\begin{figure}
    \centering
    \includegraphics[width=1\linewidth]{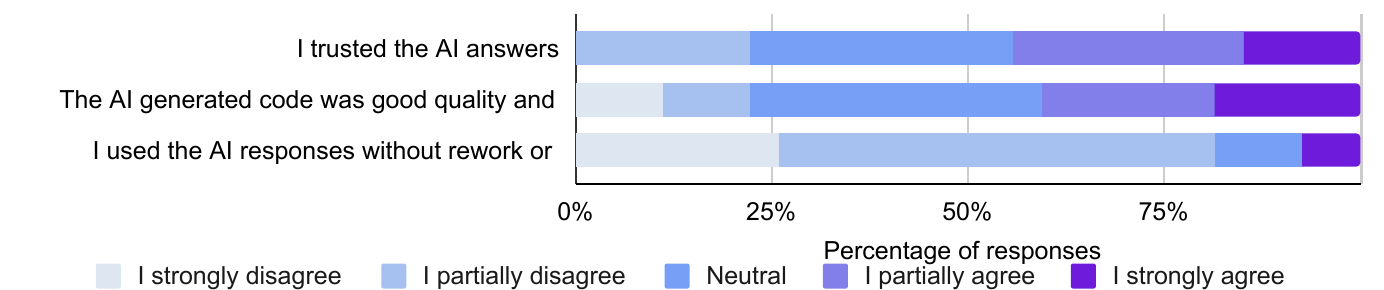}
    \caption{Quality of AI responses}
    \label{fig:quality}
    \Description[Quality of AI responses]{Shows percentages of responses that totally disagree, partially disagree, neutral, partly agree, and totally agree against three statements. I trusted the AI answers, The AI generated code was good quality, and maintainable, and I used the AI responses without rework or additional prompts.}
\end{figure}

~\autoref{fig:quality} shows respondents' perspectives on the quality of the AI responses including the amount of rework required. Trust in the answers was nearly equal between those who partially or strongly agreed to trusting the AI answers (45\%) and the remaining 55\% being neutral or partially disagreeing. Slightly fewer respondents believed the code was of good quality and maintainable (40\% partially or strongly agreed). There were stronger views on whether rework or additional prompts was required with just over 25\% of respondents strongly disagreeing with the statement that they used the responses without rework or additional prompts and 5\% strongly agreeing with the statement. Overall, the responses highlighted a general lack of trust in the quality in the code, leading to rework either by amending the code or issuing further prompts.

In considering the quality of their resulting application, only
three respondents thought it had no bugs, while two thought it
had severe bugs. The majority (22) noted it had some issues with 12
believing it had a small number, and 10 noting a moderate number.

Overall, the majority of respondents (19) felt their resulting application functional with 18 stating it was mostly functional and 1 considering it was completely functional. In contrast, 2 thought it was not functional and 6 finding it partially functional. 

These findings show that despite the short-time frame of the vibe hack, and additional rework required in terms of modifying code or revising prompts, nearly all teams were able to build a functional prototype albeit containing some bugs.

\subsubsection{Software Architecture.} 
\begin{figure}
    \centering
    \includegraphics[width=1\linewidth]{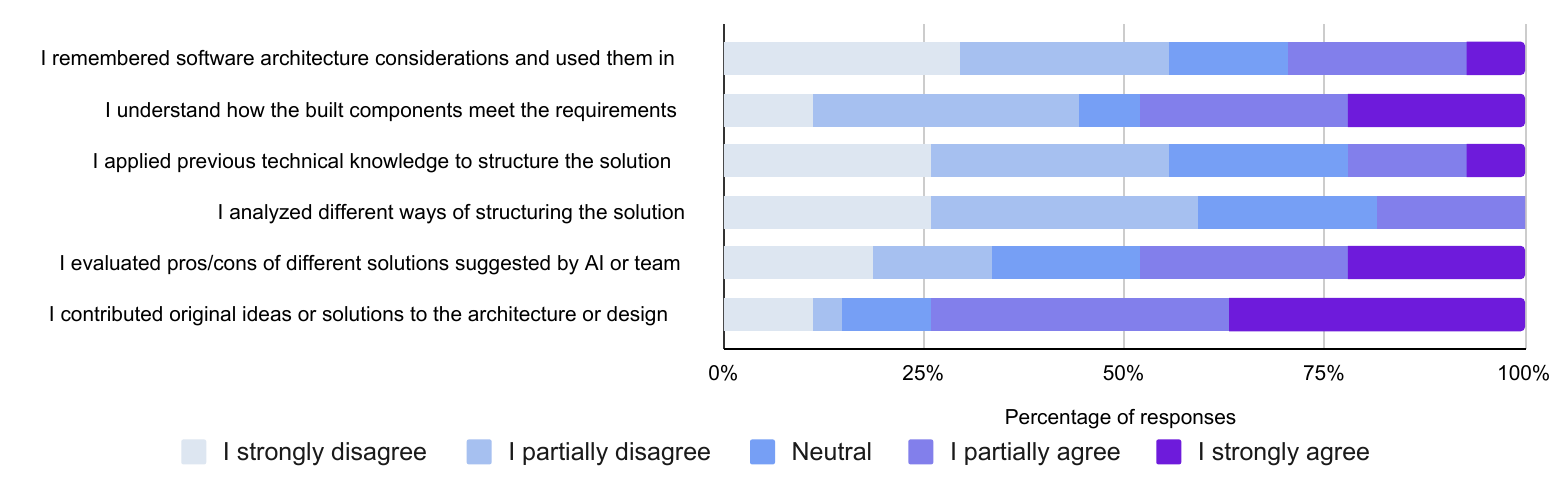}
    \caption{Architectural considerations}
    \label{fig:architecture}
    \Description[Architectural considerations]{Shows percentages of responses that totally disagree, partially disagree, neutral, partly agree, and totally agree against six  statements. I remembered software architecture considerations and used them in the project. I understand how the built components meet the requirements, I applied previous technical knowledge to structure the solution, I analyzed different ways of structuring the solution, I evaluated pros and cons of different solutions suggested by AI or team mates, and U contributed original ideas or solutions to the architecture or design.}
\end{figure}

Despite there being little knowledge of software architecture among the respondents (\autoref{prior_experience}), there was some contrasting evidence of architecture practices being used when developing the solution (see~\autoref{fig:architecture}). On the one hand, while nearly 75\% of respondents partially or fully agreed with the statement that they contributed original ideas or solutions to the architecture or design, 
fewer than 25\% of respondents partially or strongly agreed that they applied previous technical knowledge to structure the solution with even fewer agreeing they analyzed different ways of structuring the solution. This shows a contrast between architecture self-efficacy (little application of knowledge or analysis of solutions) against a perceived large number of ideas contributed.


\vicki{reworked slightly based on Kiev's arguments. Needs reviewing.}
\kiev{I have a few arguments: we intended to analyze students SE self-efficacy framed by Bloom's Taxonomy, but to shorten the survey, we only covered Software Architecture. The responses contrast with the actual effort on architecture (low or no effort in discussing architecture), so this may shed some light on investigating this biased perception?}

\subsubsection{Respondents' Learnings.} 
The main learnings from the hacka-thon were in the areas of prompt engineering (23 responses), teamwork (23), and creative thinking (21). As summed up by one respondent, \say{\textit{It was a very light and fun hackathon! We were able to test several ideas, use AI in almost every step, and learn a lot in the process, all in a super collaborative atmosphere.}} Note the first learning (prompt engineering) is aligned with the many who were motivated to attend to learn about vibe coding. Only 2 respondents indicated they had not learned anything new. 

\begin{figure}
    \centering
    \includegraphics[width=1\linewidth]{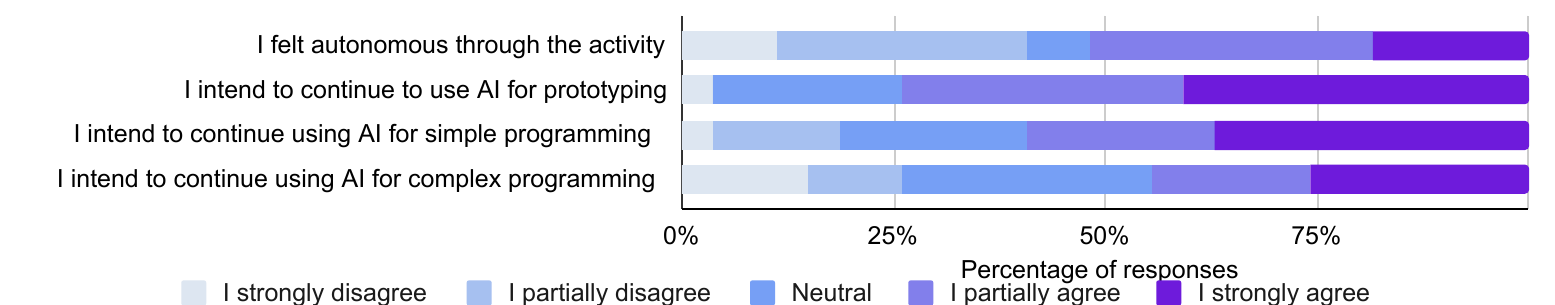}
    \caption{Intended use of AI going forward}
    \label{fig:futureAIuse}
    \Description[Intended use of AI going forward]{Shows percentages of responses that totally disagree, partially disagree, neutral, partly agree, and totally agree against four statements. I felt autonomous through the activity, I intend to continue to use AI for prototyping, I intend to continue using AI for simple programming, and I intend to continue using AI for complex programming.}
\end{figure}

In examining the intention to use AI going forward (\autoref{fig:futureAIuse}), more respondents have stronger intentions in using AI for prototyping than for programming activities. Moreover, there is stronger intent to use AI for simple programming compared to complex programming. These results perhaps highlight a view that the AI tools are strong at prototyping (including generating many ideas), but not yet good enough to develop the entire application, especially for more complex scenarios, due to quality issues leading to bugs and additional rework of the code.

\subsection{Interviews}

This section presents the interview results, with each section corresponding to the key points that emerged from the clustering process of the quotes.

\subsubsection{Intensive learning and prompt engineering as skill and outcome}

Across interviews, participants framed vibe coding as an intensive learning context where prompt engineering was both a skill and an outcome. One participant emphasized iterative experimentation with inputs and outputs, reporting that different LLMs demanded tailored prompting, \say{\textit{It was possible to train many inputs and outputs and test different input formats to get the desired output, each LLM responds to how you give the input and generates a particular output}} (P01). Others described a shift from improvisation to more structured prompting as the event progressed, \say{\textit{after some errors, we started paying more attention to prompts, writing more in each one}} (P05). Participants also learned terminology and component names through hands-on experience, as \say{\textit{you learn the names of things, what a card is, what a footer is}} (P01).

\subsubsection{Human creativity, domain knowledge, and social context}

Human creativity, domain knowledge, and design thinking were seen as necessary complements to LLMs. A recurrent pattern was that ideation benefited from facilitation and product framing. As one put it, \say{\textit{having someone skilled in design thinking to support the creative part and understand the problem pain points}} (P01). Students often attributed ownership of the idea to the team while using AI to expand options, as seen in statements such as, \say{\textit{Was it AI? No, the idea was ours from the start}} (P01) and \say{\textit{Claude raised several issues and helped validate the problem}} (P02). Social context knowledge was a critical input for the creative process, especially for LGBTQIA+ safety app, which participants said the models did not fully capture: \say{\textit{here is another social layer that ChatGPT does not grasp}} (P01).

\subsubsection{Early discovery, prototyping, and critical refinement}

Teams utilized LLMs to accelerate early discovery and prototyping, then exercised critical judgment to refine their work. Several credited AI with jump-starting brainstorming, \say{\textit{above all, AI was invaluable for the start to begin iterating}} (P02), and \say{\textit{for ideation and prototyping to validate a new idea, it is very interesting to use}} (P03). Designers highlighted how tools rapidly produced visually coherent layouts that would otherwise take longer, \say{\textit{it is as if we struggle less to reach a well-composed component}} (P07). Yet participants maintained a critical stance, adjusting CTAs, colors, and flows, \say{\textit{it will give average answers, you must read with a critical eye}} (P02).

\subsubsection{Tools comparative evaluations and combinations}

Tool comparisons were consistent. Lovable was repeatedly rated as the most effective for front-end scaffolding, while V0 drew criticism for code quality. One participant rated the outputs as follows: \say{\textit{I rated quality, Lovable 9.5 out of 10... V0 8 out of 10... to me the best tool is Lovable}} (P01), whereas another stated, \say{\textit{V0 code quality is not great, it brings odd things}} (P02). The Stitch tool was appreciated for fast prototyping but seen as mid-fidelity: \say{\textit{the screens were very mid-fidelity}} (P07). Participants frequently funneled prompts through ChatGPT to refine instructions for other generators, \say{\textit{every time we sent a prompt to Lovable or V0, it went through ChatGPT first}} (P02). The teams employed multiple tools in a pipeline-like process, leveraging their features. As P09 described: \say{\textit{I moved results from one tool to another when I needed a result that was close to what the first had given, exporting so the next tool could continue from that base}} (P09).

\subsubsection{Collaboration workflows and roles}

Collaboration practices blended human tasking with AI delegation, often mediated by prompt documents and Kanban. One team described a hybrid pipeline that separated what could be fixed by prompt from what required manual code: \say{\textit{We divided what could be solved by prompt and what required manual work}} (P01). Another approach centralized the prompts in a shared Google Docs document to reduce drift: \say{\textit{There was a shared doc where everyone built the prompt}} (P05). Some groups experimented in parallel across platforms using a prompt master: \say{\textit{We reached a prompt master, tested it across services, then refined it further}} (P09). Leadership and role fluidity emerged alongside this, such as product-oriented leadership without micromanagement, as exemplified by \say{\textit{I felt comfortable leading as the product person}} (P02), and explicit prompt engineer roles, where \say{\textit{I was responsible for prompts}} (P01).

\subsubsection{Speed gains, code quality, and stability limits}

Participants reported both speed gains and uneven code quality. Many celebrated how a single detailed prompt produced most of the front-end, noting that \say{\textit{with one prompt it generated about 80 percent of the solution}} (P03) and \say{\textit{the front was very complete and complex, alone I would take much longer}} (P05). Others flagged fragile or incoherent outputs, especially with navigation and integration, \say{\textit{in integrating screens, it failed to separate components and the structure was not good}} (P04). Stability issues were common, \say{\textit{sometimes it worked, sometimes it did not}} (P05), and participants cautioned that production standards are not met, \say{\textit{it is very hard for it [the code generation tool] to understand an application’s design patterns, even if the generated code is good [...] So, there, although it has a very good generated code, it may not follow the commit standards.}} (P03).

\subsubsection{Constraints, frictions, and coping strategies}

Constraints and frictions shaped strategy. Strict token or credit limits pushed teams toward fewer, better elaborated prompts and careful planning: \say{\textit{We only had ten [credits], so we waited to include as much as possible in one go}} (P02), and \say{\textit{We planned prompts for Lovable because we had few credits}} (P06). Some used workarounds, \say{\textit{share the project link, duplicate it, credits reset}} (P01). Time pressure also led to trusting AI too much, which backfired: \say{\textit{When time was tight, we threw things at the AI and problems started}} (P06).

\subsubsection{Negotiating AI suggestions and product scope}

Participants negotiated the boundary between human creativity and AI suggestions. Several felt large models proposed prominent features, useful for validation but not originality, \say{\textit{support to suggest some features, we wanted to go beyond the obvious}} (P01). Others appreciated average but serviceable ideas to move forward, \say{\textit{it was the average idea, which we needed for validation}} (P02). Human judgments about scope and MVP constrained AI-generated breadth, \say{\textit{we feared it would be too much, we needed the minimum viable}} (P02).

\subsubsection{Experience, background, and learning with AI}

Experience, background, and generational habits influenced how participants leveraged LLMs. Some described AI as an amplifier for novices and a tutor for coding gaps, \say{\textit{I would have been more useless without AI}} (P01), and \say{\textit{I use AI as my pet junior}} (P08). Others contrasted cohorts that learned to program pre-AI with those who began with AI at hand, \say{\textit{my programming start was entirely with AI}} (P01). Even skeptics acknowledged the value of prototyping, while warning about reliability and learning costs, noting that \say{\textit{code quality is worse and less reliable}} (P04) and that \say{\textit{prompt engineering is a skill everyone will need}} (P03).

\subsubsection{Current scope and perceived value of vibe coding}

The consensus places vibe coding’s sweet spot in early-stage design and functional prototyping, rather than production. As one summarized, \say{\textit{it exists for prototyping and needs several rounds to become a complete product}} (P09). Yet within that scope, teams reported confidence and satisfaction, \say{\textit{it was light and fun, natural language gives a sense of closeness to the solution}} (P03), and \say{\textit{in a one-day hackathon, getting to a shippable proposal is greatly helped}} (P07). Students from other fields were satisfied with learning and one of them became a vibe coding enthusiast and did a project of her own after the event: \say{\textit{I also learned a lot there ... the tools you taught. I’ve even already used them for something I was curious to do, which was a website about NLP for Language and Literature students}} (P10).

\section{Discussion}

This section synthesizes findings across our three data sources to characterize the educational experience of vibe coding hackathons and extract actionable lessons for future events.

\subsection{General Findings}

Triangulating observations, survey responses, and interviews revealed four major themes that characterize the experience of the vibe coding hackathon.

\subsubsection{Psychological safety and confidence building. } Although the exit survey rarely made this explicit, at least one open-ended response described the event as a way to gain confidence to engage with hackathons. This aligns with our qualitative evidence that generative tools can act as low-stakes scaffolds in early discovery and prototyping (e.g., participants describing faster ``first drafts'' of UIs and code), thereby reducing the perceived cost of ``getting started''. In our context, the combination of an inclusive one-day format, access to GenAI tools, and light mentoring appears to have lowered the barrier for novices without displacing human judgment or teamwork. We view this as an inclusion mechanism worth making explicit in future designs.
\vinicius{the idea here is to tie "safety/trust" to the inclusion argument}

\subsubsection{Cross-disciplinary participation as enrichment, not replacement. }
The majority of participants were computing students, but the presence of even a small number of non-computing participants (\~15\%) introduced perspectives that technical teams alone rarely highlighted. Design and humanities students reported focusing on accessibility, inclusiveness, and communication aspects of the solutions. This suggests that diversity of background—even in small proportions—can enhance the authenticity of hackathon learning outcomes. However, our data also highlight that attracting a larger non-technical population remains a challenge.

\subsubsection{Premature convergence in ideation. } 
Our survey data and open-ended comments suggest that GenAI was consistently used to rapidly turn initial ideas into concrete ones, but this often led to early convergence. While participants praised the ability of GenAI tools to generate ``first drafts'', they also tended to stop iterating once a viable idea was on the table. This pattern contrasts with the typical hackathon dynamic, where brainstorming often generates a broader range of potential solutions. Educators and organizers should therefore consider scaffolds that explicitly encourage divergent thinking (e.g., requiring multiple candidate prompts before committing), so that the benefits of GenAI do not come at the cost of creative breadth.

\subsubsection{Vibe coding as orchestration of tool pipelines. } 
\begin{figure}
    \centering
    \includegraphics[width=0.95\linewidth]{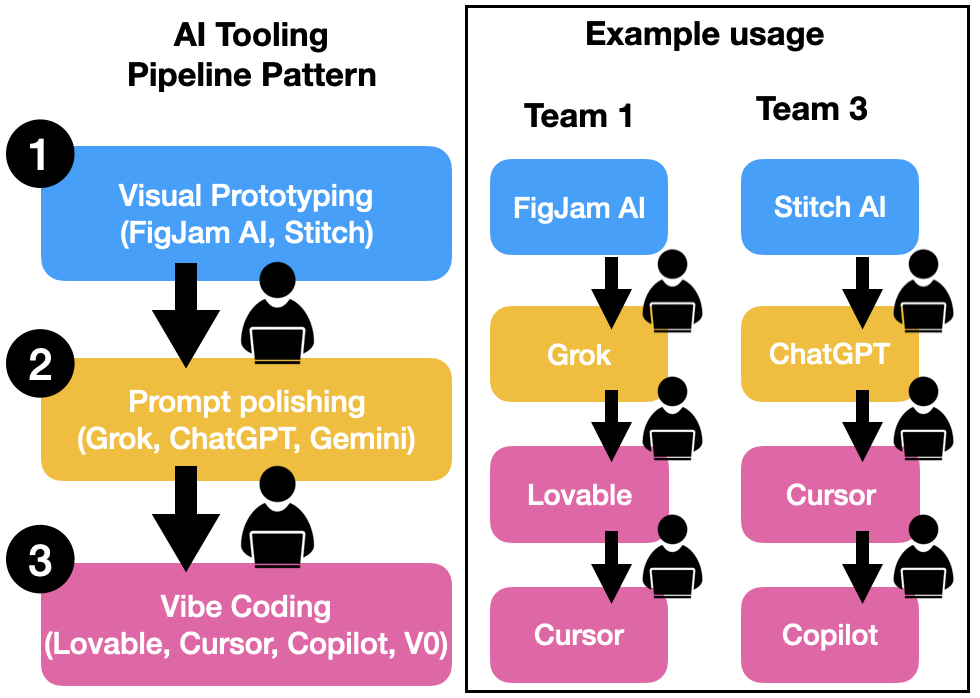}
    \caption{Typical AI tooling pipeline employed by teams}
    \label{fig:pipeline}
    \Description[Typical AI tooling pipeline employed by teams]{Shows a typical AI tooling pipeline pattern of three steps: Visual Prototyping using tools such as FigJam AI and Stitch, then Prompt polishing using tools such as ChatGPT and Gemini, and then vibe coding using tools such as Lovable, Cursor, and Copilot. Then shows example usage by two teams. Team 1 uses FigJamAI for visual prototyping, Grok for Prompt polishing, and then Lovable followed by Cursor for vibe coding. Team 3 uses Stich AI for visual prototyping, then ChatGPT for prompt polishing, and then cursor followed by Copilot for vibe coding.}
\end{figure}

For this event, there was no ``one size fits all'' tool, meaning that students combined different GenAI tools, following a recurring pipeline as illustrated in \autoref{fig:pipeline}. As the first step, they would generate prototypes with tools such as FigJam and Google Stitch. Some of these prototypes would be discarded, and the chosen one would be used as input for vibe coding tools. Usually, the prompts would be polished and the text improved with tools such as ChatGPT, Grok, or Gemini. The refined prompts would be used in tools specialized in generating apps (e.g., Lovable, Cursor, V0), and output projects would eventually be exported to be polished with other tools (e.g., Copilot).

\subsection{Lessons Learned}

This subsection synthesizes our key takeaways from the vibe-coding hackathon, grouping positive insights and areas for improvement to inform future events.

\textbf{What worked (+)}
\begin{itemize}
\item[+] \textit{Inclusive entry point:} Non-programmers were able to contribute meaningfully and, in some cases, deliver results comparable to programmers.
\item[+] \textit{Psychological safety for newcomers:} First-time participants treated the event as a confidence-building on-ramp to software creation with AI.
\item[+] \textit{Speed for prototyping:} Teams moved from ideas to functional prototypes quickly, generating and discarding alternatives at low cost.
\item[+] \textit{Prompt practices improved under constraints.} Credit and time limits encouraged careful, prompt planning, shared prompt logs, and iterative refinement.
\item[+] \textit{Cross-disciplinary value:} Even limited participation by non-computing students improved framing, accessibility, and design quality.
\item[+] \textit{Make the on-ramp explicit:} Advertising the format as low-stakes and providing a short prompt clinic and mentor access helped increase inclusivity.
\item[+] \textit{More progress in shorter time.} Enables shorter, less exhausting events, appropriate for participants with family duties and other appointments.

\end{itemize}

\textbf{What to change (-)}
\begin{itemize}
\item[-] \textit{Broaden recruitment beyond computing:} Outreach and incentives were insufficient to attract non-computing participants at scale. Future editions should partner with non-CS programs, student societies, and advisors, and should tailor messaging to emphasize creative roles, design, and domain expertise.
\item[-] \textit{Focus the challenge:} The theme generator produced heterogeneous problem scopes, which complicated evaluation and favored topics with higher perceived social impact. Provide a small menu of focused themes with aligned judging criteria, or run parallel tracks with separate awards.
\item[-] \textit{Counter premature convergence:} GenAI accelerated ideation but encouraged early lock-in. Introduce light scaffolds that promote divergence, for example, a required round with at least three distinct concepts and corresponding prompt variants before committing.
\item[-] \textit{Strengthen software engineering learning:} Production-grade practices remained limited within a one-day format. Add optional mini-checkpoints on testing, simple quality gates, and short debriefs on maintainability to connect prototyping with core SE concepts.
\end{itemize}

\subsection{Limitations and Threats to Validity}

\alex{This is a matter of style, but limitations and TTV here or in the Methods section?}

\filipe{Please, check if this description of the limitations is enough:}
\vicki{should we add something (relevant to the point above about ideation) that the students only had a short time to ideate by design and this could have impacted ideation (although there is some evidence that time pressure increases creativity so maybe the pressure is a good thing). Hmmm. What do do?}

This work has limitations, particularly since it is an experience report rather than an empirical study. Participation skewed toward computing students, which may have influenced both the pace of progress and the types of challenges encountered. The hackathon was short, which favored prototyping and limited the observation of sustained software engineering practices, including testing and refactoring. Due to a decision to keep the survey short, we did not thoroughly examine perceived SE self-efficacy, other than in the context of architecture. The study focused on a single institutional context and a specific configuration of tools, which affects transferability. We mitigated threats to validity through method triangulation, careful transcription and anonymization, and a transparent analysis process that combined open coding, model-assisted clustering, and human verification.

\section{Conclusions}



Vibe coding enabled novice and mixed-experience students to move from ideas to runnable prototypes within a single-day hackathon. Through the analysis of data from observations, survey, and interviews, we found that the approach supported rapid ideation, front-end scaffolding, and cross-disciplinary collaboration. Participants treated prompt engineering as a learnable skill, and teams developed lightweight practices to share, refine, and compare prompts.

The educational value lies in creating low-risk settings where students co-create with AI while retaining human judgment. Organizers can amplify learning by making prompting practices visible, encouraging critical review of AI outputs, and planning for practical constraints such as tool limits and integration issues. Within these boundaries, vibe coding functions well for early discovery and functional prototyping, while production-quality engineering remains out of scope.

Since this is an experience report, we do not claim generalizability. The findings are situated in a single institutional context and a short event. We offer grounded, practice-oriented takeaways for educators and hackathon organizers who wish to design inclusive activities that develop AI collaboration skills without sacrificing human oversight. Future work should examine longer formats and more diverse cohorts to connect rapid prototyping with instruction in other Software Engineering topics such as testing, architecture, and maintainability.

\section*{Acknowledgments}
ChatGPT and Claude were utilized to generate sections of this Work, including text and translations.
This work is partially supported by INES.IA (www.ines.org.br), CNPq grant 408817/2024-0.

\bibliographystyle{ACM-Reference-Format}
\bibliography{references}

@String{Computing = "Computing" }

@String{Computer = "{IEEE} Computer" }

@String{Springer = "Springer-Verlag" }

@misc{andrejkarpathy,
  type = {Tweet},
  title = {There's a New Kind of Coding I Call "Vibe Coding"},
  author = {{Andrej Karpathy [@karpathy]}},
  year = {2025},
  month = feb,
  journal = {Twitter},
  urldate = {2025-08-28},
  langid = {english},  
}

@misc{cursor,
    key = {cursor},
    title = {Cursor: The best way to code with AI},
    url = {https://cursor.com/en},
    year = {2025},
    urldate = {2025-09-23}
}

@misc{claude,
    key = {claude},
    title = {Claude},
    url = {https://claude.ai/},
    year = {2025},
    urldate = {2025-09-23}
}

@misc{replit,
    key = {replit},
    title = {Replit - Build apps and sites with AI},
    url = {https://replit.com/},
    year = {2025},
    urldate = {2025-09-23}
}

@article{krathwohlbloomrevision2002,
  title={A revision of Bloom's taxonomy: An overview},
  author={Krathwohl, David R},
  journal={Theory into practice},
  volume={41},
  number={4},
  pages={212--218},
  year={2002},
  publisher={Taylor \& Francis}
}

@article{sarkarVibeCodingProgramming2025,
  title = {Vibe Coding: Programming through Conversation with Artificial Intelligence},
  author = {Sarkar, Advait and Drosos, Ian},
  year = {2025},
  journal = {arXiv preprint arXiv:2506.23253},
  eprint = {2506.23253},
  archiveprefix = {arXiv}, 
}

@article{meske2025vibecodingreconfigurationintent,
      title={Vibe Coding as a Reconfiguration of Intent Mediation in Software Development: Definition, Implications, and Research Agenda}, 
      author={Christian Meske and Tobias Hermanns and Esther von der Weiden and Kai-Uwe Loser and Thorsten Berger},
      year={2025},
      journal={arXiv preprint arXiv:2507.21928},
      eprint={2507.21928},
      archivePrefix={arXiv},            
}

@article{geng2025exploringstudentaiinteractionsvibe,
      title={Exploring Student-AI Interactions in Vibe Coding}, 
      author={Francis Geng and Anshul Shah and Haolin Li and Nawab Mulla and Steven Swanson and Gerald Soosai Raj and Daniel Zingaro and Leo Porter},
      year={2025},
      journal={arXiv preprint arXiv:2507.22614},
      eprint={2507.22614},
      archivePrefix={arXiv},      
}

@misc{kimWhatVibeCoding2025,
  title = {What Is {{Vibe Coding}}? {{It}}'s {{Not About Turning Off Your Brain}}},
  shorttitle = {What Is {{Vibe Coding}}?},
  author = {Kim, Gene and Yegge, Steve},
  year = {2025},
  month = jun,
  journal = {IT Revolution},
  urldate = {2025-07-23},
  howpublished = {https://itrevolution.com/articles/what-is-vibe-coding-its-not-about-turning-off-your-brain/},
  langid = {american},
}

@article{oconnorVibeCodingNew2025,
  title = {`{{Vibe}} Coding' Is the New {{DIY}}},
  author = {O'Connor, Sarah},
  year = {2025},
  month = jun,
  journal = {Financial Times},
  urldate = {2025-07-23},
  chapter = {Technology}
}

@misc{rooseNotCoderAI2025,
  title = {Not a {{Coder}}? {{With A}}.{{I}}., {{Just Having}} an {{Idea Can Be Enough}}.},
  shorttitle = {Not a {{Coder}}?},
  author = {Roose, Kevin},
  year = {2025},
  month = feb,
  journal = {The New York Times},
  issn = {0362-4331},
  urldate = {2025-07-23},
  howpublished = {https://www.nytimes.com/2025/02/27/technology/personaltech/vibecoding-ai-software-programming.html},
  chapter = {Technology},
  langid = {american},
}

@misc{smithAIVibeCoding2025,
  title = {{{AI Vibe Coding}}: {{Engineers}}' {{Secret}} to {{Fast Development}} - {{IEEE Spectrum}}},
  shorttitle = {{{AI Vibe Coding}}},
  author = {Smith, Matthew S.},
  year = {2025},
  month = apr,
  urldate = {2025-07-23},
  howpublished = {https://spectrum.ieee.org/vibe-coding},
  langid = {english},
}

@misc{stokel-walkerWhatVibeCoding2025,
  title = {What Is Vibe Coding, Should You Be Doing It, and Does It Matter?},
  author = {{Stokel-Walker}, Chris},
  year = {2025},
  month = mar,
  journal = {New Scientist},
  urldate = {2025-07-23},
  howpublished = {https://www.newscientist.com/article/2473993-what-is-vibe-coding-should-you-be-doing-it-and-does-it-matter/},
  langid = {american},
}

@misc{talagalaWhatVibeCoding2025,
  title = {What {{Is Vibe Coding}}? {{And Why Should You Care}}?},
  shorttitle = {What {{Is Vibe Coding}}?},
  author = {Talagala, Nisha},
  year = {2025},
  month = mar,
  journal = {Forbes},
  urldate = {2025-07-23},
  chapter = {AI},
  howpublished = {https://www.forbes.com/sites/nishatalagala/2025/03/30/what-is-vibe-coding-and-why-should-you-care/},
  langid = {english}
}

@misc{williamsWhatVibeCoding2025,
  title = {What Is Vibe Coding, Exactly?},
  author = {Williams, Rhiannon},
  year = {2025},
  month = apr,
  journal = {MIT Technology Review},
  urldate = {2025-07-23},
  howpublished = {https://www.technologyreview.com/2025/04/16/1115135/what-is-vibe-coding-exactly/},
  langid = {english}
}

@article{pimenovagoodvibrations2025,
  title={Good Vibrations? A Qualitative Study of Co-Creation, Communication, Flow, and Trust in Vibe Coding},
  author={Pimenova, Veronica and Fakhoury, Sarah and Bird, Christian and Storey, Margaret-Anne and Endres, Madeline},
  journal={arXiv preprint arXiv:2509.12491},
  year={2025}
}

@inproceedings{pratherWideningGap2024,
    author = {Prather, James and Reeves, Brent N and Leinonen, Juho and MacNeil, Stephen and Randrianasolo, Arisoa S and Becker, Brett A. and Kimmel, Bailey and Wright, Jared and Briggs, Ben},
    title = {The Widening Gap: The Benefits and Harms of Generative AI for Novice Programmers},
    year = {2024},
    isbn = {9798400704758},
    publisher = {Association for Computing Machinery},
    address = {New York, NY, USA},
    url = {https://doi.org/10.1145/3632620.3671116},
    doi = {10.1145/3632620.3671116}, 
    booktitle = {Proceedings of the 2024 ACM Conference on International Computing Education Research - Volume 1},
    pages = {469–486},
    numpages = {18},
    location = {Melbourne, VIC, Australia},
    series = {ICER '24}
}

@article{yilmazGenAIEfficacy2023,
title = {The effect of generative artificial intelligence (AI)-based tool use on students' computational thinking skills, programming self-efficacy and motivation},
journal = {Computers and Education: Artificial Intelligence},
volume = {4},
pages = {100147},
year = {2023},
issn = {2666-920X},
doi = {https://doi.org/10.1016/j.caeai.2023.100147},
url = {https://www.sciencedirect.com/science/article/pii/S2666920X23000267},
author = {Ramazan Yilmaz and Fatma Gizem {Karaoglan Yilmaz}},
}

@inproceedings{keuningStudentPerceptions,
    author = {Keuning, Hieke and Alpizar-Chacon, Isaac and Lykourentzou, Ioanna and Beehler, Lauren and K\"{o}ppe, Christian and de Jong, Imke and Sosnovsky, Sergey},
    title = {Students' Perceptions and Use of Generative AI Tools for Programming Across Different Computing Courses},
    year = {2024},
    isbn = {9798400710384},
    publisher = {Association for Computing Machinery},
    address = {New York, NY, USA},
    url = {https://doi.org/10.1145/3699538.3699546},
    doi = {10.1145/3699538.3699546},    
    booktitle = {Proceedings of the 24th Koli Calling International Conference on Computing Education Research},
    articleno = {14},
    numpages = {12},    
    series = {Koli Calling '24}
}

@article{prather2023s,
  title={“It’s weird that it knows what i want”: Usability and interactions with copilot for novice programmers},
  author={Prather, James and Reeves, Brent N and Denny, Paul and Becker, Brett A and Leinonen, Juho and Luxton-Reilly, Andrew and Powell, Garrett and Finnie-Ansley, James and Santos, Eddie Antonio},
  journal={ACM transactions on computer-human interaction},
  volume={31},
  number={1},
  pages={1--31},
  year={2023},
  publisher={ACM New York, NY}
}

@inproceedings{tuzmenUseVV2024,
  title={Use of Generative Artificial Intelligence in the Education of Software Verification and Validation},
  author={Tuzmen, Ayca},
  booktitle={2024 IEEE Frontiers in Education Conference (FIE)},
  pages={1--6},
  year={2024},
  organization={IEEE},
  publisher={IEEE},
}

@inproceedings{petrovskaIncorporateGenAI2024,
author = {Petrovska, Olga and Clift, Lee and Moller, Faron and Pearsall, Rebecca},
title = {Incorporating Generative AI into Software Development Education},
year = {2024},
isbn = {9798400709326},
publisher = {Association for Computing Machinery},
address = {New York, NY, USA},
url = {https://doi.org/10.1145/3633053.3633057},
doi = {10.1145/3633053.3633057},
booktitle = {Proceedings of the 8th Conference on Computing Education Practice},
pages = {37–40},
numpages = {4},
location = {Durham, United Kingdom},
series = {CEP '24}
}

@inproceedings{houEroding2025,
author = {Hou, Irene and Man, Owen and Hamilton, Kate and Muthusekaran, Srishty and Johnykutty, Jeffin and Zadeh, Leili and MacNeil, Stephen},
title = { 'All Roads Lead to ChatGPT': How Generative AI is Eroding Social Interactions and Student Learning Communities},
year = {2025},
isbn = {9798400715679},
publisher = {Association for Computing Machinery},
address = {New York, NY, USA},
url = {https://doi.org/10.1145/3724363.3729024},
doi = {10.1145/3724363.3729024},
booktitle = {Proceedings of the 30th ACM Conference on Innovation and Technology in Computer Science Education V. 1},
pages = {79–85},
numpages = {7},
location = {Nijmegen, Netherlands},
series = {ITiCSE 2025}
}

@inproceedings{kerslakePrompting2024,
    author = {Kerslake, Chris and Denny, Paul and Smith, David H., IV and Prather, James and Leinonen, Juho and Luxton-Reilly, Andrew and MacNeil, Stephen},
    title = {Integrating Natural Language Prompting Tasks in Introductory Programming Courses},
    year = {2024},
    isbn = {9798400705984},
    publisher = {Association for Computing Machinery},
    address = {New York, NY, USA},
    url = {https://doi.org/10.1145/3649165.3690125},
    doi = {10.1145/3649165.3690125},    
    booktitle = {Proceedings of the 2024 on ACM Virtual Global Computing Education Conference V. 1},
    pages = {88–94},
    numpages = {7},
    location = {Virtual Event, NC, USA},
    series = {SIGCSE Virtual 2024}
}

@article{camaraGenerativeModeling2024,
  title={Generative ai in the software modeling classroom: An experience report with chatgpt and unified modeling language},
  author={C{\'a}mara, Javier and Troya, Javier and Montes-Torres, Julio and Jaime, Francisco J},
  journal={IEEE Software},
  volume={41},
  number={6},
  pages={73--81},
  year={2024},
  publisher={IEEE}
}

@inproceedings{pratherBeyondTheHype2025,
    author = {Prather, James and Leinonen, Juho and Kiesler, Natalie and Gorson Benario, Jamie and Lau, Sam and MacNeil, Stephen and Norouzi, Narges and Opel, Simone and Pettit, Vee and Porter, Leo and Reeves, Brent N. and Savelka, Jaromir and Smith, David H., IV and Strickroth, Sven and Zingaro, Daniel},
    title = {Beyond the Hype: A Comprehensive Review of Current Trends in Generative AI Research, Teaching Practices, and Tools},
    year = {2025},
    isbn = {9798400712081},
    publisher = {Association for Computing Machinery},
    address = {New York, NY, USA},
    url = {https://doi.org/10.1145/3689187.3709614},
    doi = {10.1145/3689187.3709614},    
    booktitle = {2024 Working Group Reports on Innovation and Technology in Computer Science Education},
    pages = {300–338},
    numpages = {39},
    location = {Milan, Italy},
    series = {ITiCSE 2024}
}

@misc{stackoverflow_dev_survey_2025,
        key={Stack Overflow},
	title = {2025 {Stack} {Overflow} {Developer} {Survey}},
	url = {https://survey.stackoverflow.co/2025},
	language = {en},
        year = {2025},
	urldate = {2025-01-29},
}

@INPROCEEDINGS{alkamliEthicalLLM2024,
  author={Alkamli, Shahad and Al-Yahya, Maha and Alyahya, Khulood},
  booktitle={2024 2nd International Conference on Foundation and Large Language Models (FLLM)}, 
  title={Ethical and Legal Considerations of Large Language Models: A Systematic Review of the Literature}, 
  year={2024},
  volume={},
  number={},
  pages={576-586},
    publisher={IEEE},
  doi={10.1109/FLLM63129.2024.10852451}}

@article{huangbias2024,
  title={Bias testing and mitigation in llm-based code generation},
  author={Huang, Dong and Zhang, Jie M and Bu, Qingwen and Xie, Xiaofei and Chen, Junjie and Cui, Heming},
  journal={ACM Transactions on Software Engineering and Methodology},
  year={2024},
  publisher={ACM New York, NY}
}

@article{falkFutureHackathonResearch2024,
  title = {The Future of Hackathon Research and Practice},
  author = {Falk, Jeanette and Nolte, Alexander and Huppenkothen, Daniela and Weinzierl, Marion and Gama, Kiev and Spikol, Daniel and Tollerud, Erik and Hong, Neil P. Chue and Kn{\"a}pper, Ines and Hayden, Linda Bailey},
  year = {2024},
  journal = {IEEE access : practical innovations, open solutions},
  volume = {12},
  pages = {133406--133425},
  doi = {10.1109/ACCESS.2024.3455092},

}

@article{nolte2020support,
  title={How to support newcomers in scientific hackathons-an action research study on expert mentoring},
  author={Nolte, Alexander and Hayden, Linda Bailey and Herbsleb, James D},
  journal={Proceedings of the ACM on Human-Computer Interaction},
  volume={4},
  number={CSCW1},
  pages={1--23},
  year={2020},
  publisher={ACM New York, NY, USA}
}

@article{gama2023developers,
  title={The developers’ design thinking toolbox in hackathons: a study on the recurring design methods in software development marathons},
  author={Gama, Kiev and Valen{\c{c}}a, George and Alessio, Pedro and Formiga, Rafael and Neves, Andr{\'e} and Lacerda, Nycolas},
  journal={International Journal of Human--Computer Interaction},
  volume={39},
  number={12},
  pages={2269--2291},
  year={2023},
  publisher={Taylor \& Francis}
}

@inproceedings{taylor2018everybody,
  title={Everybody's hacking: Participation and the mainstreaming of hackathons},
  author={Taylor, Nick and Clarke, Loraine},
  booktitle={Proceedings of the 2018 CHI Conference on Human Factors in Computing Systems},
  pages={1--12},
  year={2018},
  publisher={Association for Computing Machinery},
}

@article{pe2022corporate,
  title={Corporate hackathons, how and why? A multiple case study of motivation, projects proposal and selection, goal setting, coordination, and outcomes},
  author={Pe-Than, Ei Pa Pa and Nolte, Alexander and Filippova, Anna and Bird, Christian and Scallen, Steve and Herbsleb, James},
  journal={Human--Computer Interaction},
  volume={37},
  number={4},
  pages={281--313},
  year={2022},
  publisher={Taylor \& Francis}
}

@inproceedings{nolte2019touched,
  title={Touched by the hackathon: a study on the connection between hackathon participants and start-up founders},
  author={Nolte, Alexander},
  booktitle={Proceedings of the 2nd ACM SIGSOFT international workshop on software-intensive business: start-ups, platforms, and ecosystems},
  pages={31--36},
  year={2019}
}

@article{johnson2014civic,
  title={Civic hackathons: Innovation, procurement, or civic engagement?},
  author={Johnson, Peter and Robinson, Pamela},
  journal={Review of policy research},
  volume={31},
  number={4},
  pages={349--357},
  year={2014},
  publisher={Wiley Online Library}
}

@article{yuan2021open,
  title={Open innovation in the public sector: creating public value through civic hackathons},
  author={Yuan, Qianli and Gasco-Hernandez, Mila},
  journal={Public Management Review},
  volume={23},
  number={4},
  pages={523--544},
  year={2021},
  publisher={Taylor \& Francis}
}

@INPROCEEDINGS{gama2017preliminary,
  author={Gama, Kiev},
  booktitle={2017 IEEE/ACM 4th International Workshop on CrowdSourcing in Software Engineering (CSI-SE)}, 
  title={Preliminary Findings on Software Engineering Practices in Civic Hackathons}, 
  year={2017},
  volume={},
  number={},
  pages={14-20},
    publisher = {IEEE},
  doi={10.1109/CSI-SE.2017.5}}

@inproceedings{gamaHackathonsFormalLearning2018,
  title = {Hackathons in the Formal Learning Process},
  booktitle = {Proceedings of the 23rd Annual {{ACM}} Conference on Innovation and Technology in Computer Science Education},
  author = {Gama, Kiev and Alencar Gon{\c c}alves, Breno and Alessio, Pedro},
  year = {2018},
  series = {{{ITiCSE}} 2018},
  pages = {248--253},
  publisher = {Association for Computing Machinery},
  address = {Larnaca, Cyprus and New York, NY, USA},
  doi = {10.1145/3197091.3197138},
  isbn = {978-1-4503-5707-4},
  }

@article{affia2025organize,
  title={How to organize an in-person, online or hybrid hackathon--A revised planning kit},
  author={Affia, Abasi-Amefon Obot and Gama, Kiev and Herbsleb, James D and Nolte, Alexander},
  journal={arXiv preprint arXiv:2008.08025},
  year={2025}
}

@article{willis2017challenge,
  title={Challenge-Based Learning},
  author={Willis, Scooter and Byrd, Greg and Johnson, Brian David},
  journal={Computer},
  volume={50},
  number={7},
  pages={13--16},
  year={2017},
  publisher={IEEE}
}

@inproceedings{warner2017hack,
  title={Hack. edu: Examining how college hackathons are perceived by student attendees and non-attendees},
  author={Warner, Jeremy and Guo, Philip J},
  booktitle={Proceedings of the 2017 ACM Conference on International Computing Education Research},
  pages={254--262},
  year={2017},
  publisher={Association for Computing Machinery},
}

@inproceedings{nandi2016hackathons,
  title={Hackathons as an informal learning platform},
  author={Nandi, Arnab and Mandernach, Meris},
  booktitle={Proceedings of the 47th ACM Technical Symposium on Computing Science Education},
  pages={346--351},
  year={2016},
  organization={ACM},
  publisher={Association for Computing Machinery},
}

@inproceedings{hogan2022hackathons,
  title={Hackathons as a Tool for Authentic Learning},
  author={Hogan, Mair{\'e}ad},
  booktitle={Proceedings of the 27th ACM Conference on on Innovation and Technology in Computer Science Education Vol. 2},
  pages={582--584},
  publisher={Association for Computing Machinery},
  year={2022}
}

@inproceedings{steglich2020hackathons,
  title={Hackathons as a pedagogical strategy to engage students to learn and to adopt software engineering practices},
  author={Steglich, Caio and Salerno, Larissa and Fernandes, Tha{\'\i}s and Marczak, Sabrina and Dutra, Alessandra and Bacelo, Ana Paula and Trindade, C{\'a}ssio},
  booktitle={Proceedings of the XXXIV Brazilian Symposium on Software Engineering},
  pages={670--679},
  year={2020},
  publisher={Association for Computing Machinery},
}

@inproceedings{paganini2020engaging,
  title={Engaging Women’s Participation in Hackathons: A Qualitative Study with Participants of a Female-focused Hackathon},
  author={Paganini, Lavinia and Gama, Kiev},
  booktitle={International Conference on Game Jams, Hackathons and Game Creation Events 2020},
  pages={8--15},
  year={2020},
  publisher={Association for Computing Machinery},
}

@article{crusoe2016channeling,
  title={Channeling Community Contributions to Scientific Software: A sprint Experience},
  author={Crusoe, Michael R and Brown, C Titus},
  journal={Journal of open research software},
  volume={4},
  number={1},
  year={2016},
  publisher={NIH Public Access}
}

@inproceedings{amefon2022,
author = {Affia, Abasi-amefon Obot and Nolte, Alexander and Matulevi\v{c}ius, Raimundas},
title = {Integrating Hackathons into an Online Cybersecurity Course},
year = {2022},
isbn = {9781450392259},
publisher = {Association for Computing Machinery},
address = {New York, NY, USA},
doi = {10.1145/3510456.3514151},
booktitle = {Proceedings of the ACM/IEEE 44th International Conference on Software Engineering: Software Engineering Education and Training},
pages = {134–145},
numpages = {12},
location = {Pittsburgh, Pennsylvania},
series = {ICSE-SEET '22}
}

@inproceedings{steglich2021online,
  title={An Online Educational Hackathon to Foster Professional Skills and Intense Collaboration on Software Engineering Students},
  author={Steglich, Caio and Marczak, Sabrina and Guerra, Luiz and Trindade, C{\'a}ssio and Dutra, Alessandra and Bacelo, Ana},
  booktitle={Brazilian Symposium on Software Engineering},
  pages={388--397},
  year={2021},
  publisher={Association for Computing Machinery},
}

@inproceedings{porras2018hackathons,
  title={Hackathons in software engineering education: lessons learned from a decade of events},
  author={Porras, Jari and Khakurel, Jayden and Ikonen, Jouni and Happonen, Ari and Knutas, Antti and Herala, Antti and Dr{\"o}gehorn, Olaf},
  booktitle={Proceedings of the 2nd intl workshop on software engineering education for Millennials},
  pages={40--47},
  publisher={Association for Computing Machinery},
  year={2018}
}

@inproceedings{falk202010,
  title={10 years of research with and on hackathons},
  author={Falk Olesen, Jeanette and Halskov, Kim},
  booktitle={Proceedings of the 2020 ACM designing interactive systems conference},
  pages={1073--1088},
  publisher = {Association for Computing Machinery},
  year={2020}
}

@misc{supplementarymaterial,
    author={Anonymous},
    title ={Supplementary Material},
    howpublished = {\url{https://figshare.com/s/cccbd87e316db0079e29}},
    year = {2025},
}

@incollection{resnick2009growing,
  title={Growing up programming: democratizing the creation of dynamic, interactive media},
  author={Resnick, Mitchel and Flanagan, Mary and Kelleher, Caitlin and MacLaurin, Matthew and Ohshima, Yoshiki and Perlin, Ken and Torres, Robert},
  booktitle={CHI'09 Extended Abstracts on Human Factors in Computing Systems},
  pages={3293--3296},
  year={2009},
  publisher={Association for Computing Machinery},
}

@article{robins2003learning,
  title={Learning and teaching programming: A review and discussion},
  author={Robins, Anthony and Rountree, Janet and Rountree, Nathan},
  journal={Computer science education},
  volume={13},
  number={2},
  pages={137--172},
  year={2003},
  publisher={Taylor \& Francis}
}

@article{kara2019challenges,
  title={Challenges faced by adult learners in online distance education: A literature review},
  author={Kara, Mehmet and Erdogdu, Fatih and Koko{\c{c}}, Mehmet and Cagiltay, Kursat},
  journal={Open Praxis},
  volume={11},
  number={1},
  pages={5--22},
  year={2019}
}

@article{freeman2014active,
  title={Active learning increases student performance in science, engineering, and mathematics},
  author={Freeman, Scott and Eddy, Sarah L and McDonough, Miles and Smith, Michelle K and Okoroafor, Nnadozie and Jordt, Hannah and Wenderoth, Mary Pat},
  journal={Proceedings of the national academy of sciences},
  volume={111},
  number={23},
  pages={8410--8415},
  year={2014},
  publisher={National Academy of Sciences}
}

@inproceedings{trinkenreich2025get,
  title={Get on the Train or be Left on the Station: Using LLMs for Software Engineering Research},
  author={Trinkenreich, Bianca and Calefato, Fabio and Hanssen, Geir and Blincoe, Kelly and Kalinowski, Marcos and Pezz{\`e}, Mauro and Tell, Paolo and Storey, Margaret-Anne},
  booktitle={Proceedings of the 33rd ACM International Conference on the Foundations of Software Engineering},
  pages={1503--1507},
  year={2025}
}

@article{bano2024large,
  title={Large language models for qualitative research in software engineering: exploring opportunities and challenges},
  author={Bano, Muneera and Hoda, Rashina and Zowghi, Didar and Treude, Christoph},
  journal={Automated Software Engineering},
  volume={31},
  number={1},
  pages={8},
  year={2024},
  publisher={Springer}
}

\end{document}